\documentstyle[12pt]{article}
\begin{document}
{\hskip 12.cm} SNUTP 97-034\par
\vspace{1ex}
{\hskip 12.0cm} September, 1997\\
\vspace{5ex}
\begin{center}        
{\LARGE \bf $1/N_c$ Expansion of the Heavy Baryon Isgur-Wise Functions}\\
\vspace{7ex}
{\sc Chun Liu}\footnote{\it email: liuc@ctp.snu.ac.kr}\\
\vspace{3ex}     
{\it Center For Theoretical Physics, Seoul National University}\\
{\it Seoul, 151-742, Korea}\\

\vspace{10.0ex}
{\large \bf Abstract}\\
\vspace{4ex}
\begin{minipage}{130mm}
   
   The $1/N_c$ expansion of the heavy baryon Isgur-Wise functions is 
discussed.  Because of the contracted 
$SU(2N_f)$ light quark spin-flavor symmetry, the universality relations
among the Isgur-Wise functions of $\Lambda_b\rightarrow \Lambda_c$ and
$\Sigma_b^{(*)}\rightarrow \Sigma_c^{(*)}$ are valid up to the order of
$1/N_c^2$. \\

\vspace{2.0cm}
\par
\end{minipage}
\end{center}

\newpage
                                                                               
   Heavy baryons provide us with a testing ground for the Standard Model 
(SM), especially to Quantum Chromodynamics (QCD) in some aspects.  With 
the accumulation of the experimental data on heavy baryons, some important
parameters of SM, for instance the Cabbibo-Kobayashi-Maskawa (CKM) matrix 
element $V_{cb}$, can be extracted by comparing experiments with 
theoretical calculations.  The main difficulties in the calculations are 
due to our poor understanding of the nonperturbative QCD.  In this Brief
Report, we discuss the $1/N_c$ expansion [1] for the 
heavy baryon weak decay form factors.  We will point out that it
can be applied to relate different baryon Isgur-Wise functions with  
a comparative accuracy.  
\par
\vspace{1.0cm}    
   Heavy baryon weak decays can be systematically studied by the heavy quark 
effective theory (HQET) [2].  The classification of the form factors 
parameterizing the hadronic matrix elements of the weak currents is 
simplified significantly [3].  Under the heavy quark limit, only one universal 
form factor remains to be determined in the $\Lambda_b\rightarrow \Lambda_c$
transition, and two in $\Sigma_b^{(*)}\rightarrow \Sigma_c^{(*)}$ transitions.
These universal
form factors are called Isgur-Wise functions. 
They should be calculated by some 
nonperturbative method.
\par
\vspace{1.0cm}
   Large $N_c$ limit is one of the most important and model-independent 
method
of nonperturbative QCD.  Nonperturbative properties of mesons can be observed 
from the analysis of the planar diagram, and baryons from the Hartree-Fock
picture.  Recently there are renewed interests in the large $N_c$
application to baryons [4-7].  It is pointed out that there
is a contracted SU(2$N_f$) light quark spin-flavor symmetry in the baryon
sector in the large $N_c$ limit.  The observation of this light quark 
spin-flavor symmetry results in many quantitative applications [8-14]. 
In the large $N_c$ limit, the relations among the baryon Isgur-Wise functions 
have been studied [11-13].  With the definitions of 
\begin{equation}
\begin{array}{lll}
<\Lambda_c(v',s')|\bar{c}\Gamma b|\Lambda_b(v,s)>&=&
\eta(y)\bar{u}_{\Lambda_c}(v',s')\Gamma u_{\Lambda_b}(v,s)~,\\[3mm]
<\Sigma_c^{(*)}(v',s')|\bar{c}\Gamma b|\Sigma_b^{(*)}(v,s)>&=&
[\xi(y)g_{\mu\nu}+\zeta(y)v_{\nu}v'_{\mu}]
\bar{u}_{\Sigma_c^{(*)}}^{\nu}(v',s')\Gamma u_{\Sigma_b^{(*)}}^{\mu}(v,s)~,
\end{array}
\end{equation}
where $y=v\cdot v'$, $u_{\Sigma_Q^*}^{\mu}(v,s)$ is the Rarita-Schwinger 
spinor and $u_{\Sigma_Q}^{\mu}(v,s)$ is defined by 
\begin{equation}
u_{\Sigma_Q}^{\mu}(v,s)=\frac{(\gamma^{\mu}+v^{\mu})\gamma_5}{\sqrt{3}}
u_{\Sigma_Q}(v,s)~,
\end{equation}
the Isgur-Wise functions $\eta(y)$, $\xi(y)$ and $\zeta(y)$ have the 
following large $N_c$ relations [11, 12],
\begin{equation}
\eta(y)=\xi(y)=-(y+1)\zeta(y)~.
\end{equation}
\par
\vspace{1.0cm}
   We study the $1/N_c$ expansion of the relations among the baryon Isgur-Wise 
functions.  It is interesting because of the observation of the light quark
spin-flavor symmetry in the baryon sector in the large $N_c$ limit.
It is heuristic to illustrate the large $N_c$ baryon
Isgur-Wise functions in a naive way which can be directly applied to 
the discussions of the $1/N_c$ corrections.  Under the heavy quark limit,
the heavy quark in the heavy hadron has fixed velocity which is identical 
with the heavy hadron velocity.  And the heavy quark spin decouples from its 
strong interaction with the light quark system in the hadron.  
The light quark system cannot see any properties of the heavy quark except 
its velocity.  In this case, the spin $J_l$ and isospin $I_l$ of the light
quark system become good quantum numbers to describe the heavy baryons in 
which, quarks have no orbital angular momentum excitations in the constituent 
picture, and there are only two flavors of light quarks.  For the number of 
colors being $N_c$, the baryons we are interested in are  
$(I_l, J_l) = (0, 0), (1, 1), ..., (\frac{N_c-1}{2}, \frac{N_c-1}{2})$.
When the velocity of the heavy quark changes from $v$ to $v'$ due to weak 
decay, the brown muck has to undergo a transition through the strong 
interaction from the heavy quark.  The Isgur-Wise functions defined in Eqs. 
(1) and (2) just 
measure the amplitudes of the brown muck transfers.  They cannot be determined 
from the HQET further, however.  It is at this stage, the large $N_c$ method
is applied.  In the large $N_c$ limit, there is the SU(4) light quark
spin-flavor symmetry for baryons [4].  During the transition, the spin of 
any light quark in the brown muck is conserved.  In other words, the light 
quark spin individually decouples from the strong interaction in the brown
muck transition.  The Isgur-Wise functions are independent of the 
light quark spin configuration in the brown muck in the large $N_c$ limit, 
and therefore deserve an SU(4) expansion.  At the leading order of the SU(4)
expansion, the relations given by Eq. (3) still hold.   
This is the 
essential point in deriving the large $N_c$ universality of the baryon 
Isgur-Wise functions for the $\Lambda_b\rightarrow \Lambda_c$ and
$\Sigma_b^{(*)}\rightarrow \Sigma_c^{(*)}$ decays in Ref. [12] based on the 
SU(4) symmetry.  
\par
\vspace{1.0cm}
   Let us go further to the corrections of this contracted SU(4) light quark
spin-flavor symmetry result.  
They can be considered straightforwardly 
along above discussion.  The corrections only appear in the finite $N_c$
case.  
We note the baryon 
spectrum has the relation $I_l=J_l$.  Therefore the Isgur-Wise functions 
have the following expansions,
\begin{equation}
\begin{array}{lll}
\eta(y) &=&\displaystyle\tilde{\eta}_0(y)~,\\[3mm]
\xi(y)  &=&\displaystyle\tilde{\eta}_0(y)+\tilde{\xi}(y)
\frac{J_l^2}{N_c^2}~,\\[3mm]
-(y+1)\zeta(y)&=&\displaystyle\tilde{\eta}_0(y)+\tilde{\zeta}(y)\frac{J_l^2}
{N_c^2}~,\\[3mm]
\end{array}
\end{equation} 
where $\tilde{\eta}_0(y)$ is the leading SU(4) symmetry result which is 
independent of the brown muck spin or isospin.  $\tilde{\xi}(y)$ and 
$\tilde{\zeta}(y)$ parameterize the SU(4) breaking effects, and are normalized
to be order $1$ in the large $N_c$ limit.  
The factor $N_c^2$ should be there so
as to keep the $N_c$ scaling for the Isgur-Wise functions.  In the extreme
case while in the baryon all the light quark spins align in the same 
direction, $J_l^2$ scales as $N_c^2/4$.  Only by dividing a factor $N_c^2$,
have the terms proportional to $J_l^2$ in above equation the right $N_c$ 
scaling.  Note there is no term which has linear dependence on $J_l$ in the
corrections.  This is simply because there is no way to combine $J_l$ with
$(v-v')^{\mu}$ into a Lorentz and CP invariant quantity.  
Generally, the Isgur-Wise function should depend on $J_l/N_c$ or 
$I_l/N_c$ of the brown muck undergoing the transition.
However it is interesting to note that the Isgur-Wise function $\eta(y)$ 
does not have any corrections in the SU(4) expansion because $\Lambda_Q$ 
baryon is a SU(4) singlet.  From Eq. (4), we see that the SU(4) symmetry 
relations (3) are valid up to the order of $1/N_c^2$.  
\par
\vspace{1.0cm}
   It is helpful for the understanding to compare this framework with the
heavy quark Skyrme model [15, 16] which is often believed to be the large
$N_c$ HQET.  The model predicted an exponential form for the Isgur-Wise 
function $\eta(y)$ [17].  But the way to calculate its $1/N_c$ corrections
is not available.  On the other hand, as is well-known, 
there are $O(1/N_c)$ contributions
in $\tilde{\eta}_0(y)$, we have no knowledge about $\tilde{\eta}_0(y)$ itself,
however the predicted universality relations of the Isgur-Wise function, 
Eq. (4), are valid up to the order of $1/N_c^2$.  (The analogous situation
holds for the case of baryon masses [9, 14].) If we focus on the relation 
among the Isgur-Wise functions, the SU(4) expansion provides a better 
framework.  In the near future, the 
Isgur-Wise function $\eta(y)$ can be extracted from the experimental data 
of the $\Lambda_b\rightarrow \Lambda_c$ semileptonic decay.  With this 
information, the weak decays $\Omega_b^{(*)}\rightarrow \Omega_c^{(*)}$ can
be predicted to a comparatively accurate level in the chiral SU(3) symmetry 
limit. 
\par
\vspace{1.0cm}
   Finally let us add a remark on the $1/m_Q$ expansion for the form factors.
HQET is a systematic method for the expansion [18].  What we have discussed in 
this note is just the leading order of heavy quark expansion.  To the order of 
$1/m_Q$, there is an additional universal form factor for the 
$\Lambda_b\rightarrow \Lambda_c$ semileptonic decay.  This form factor 
vanishes in the limit of $N_c\rightarrow\infty$.  Therefore when we extract
$\eta(y)$ from the decay of $\Lambda_b\rightarrow \Lambda_c$, it subjects
to an uncertainty of $\Lambda_{\it QCD}/N_cm_Q$.  This uncertainty can be the
same order of $1/N_c^2$ numerically, therefore does not spoil the accuracy we
hoped to achieve.
\par
\vspace{2.0cm}

   I am grateful to Mark Wise for helpful discussions.  This work was 
supported by the Korea Science and Engineering Foundation
through the SRC program.

\end{document}